# Quantum simulation of single-server Markovian queues: A dynamic amplification approach

Michal Koren[1*], Or Peretz[1]

[1] School of Industrial Engineering and Management, Shenkar—Engineering. Design. Art, Ramat-Gan, Israel

*Contact author: michal.koren@shenkar.ac.il

Michal Koren, email: michal.koren@shenkar.ac.il

Or Peretz, email: or.perets@shenkar.ac.il

Corresponding author:
Michal Koren
Shenkar—Engineering. Design. Art.
School of Industrial Engineering and Management
Anne Frank 12, Ramat Gan 5252317, Israel

**ABSTRACT**

Quantum computing is revolutionizing various fields, including operations research and queueing theory. This study presents a quantum method for simulating single-server Markovian (M/M/1) queues, making quantum computing more accessible to researchers in operations research. We introduce a dynamic amplification approach that adapts to queue traffic, potentially improving simulation efficiency, and design custom-parameterized quantum gates for arrival and service processes. This flexible framework enables modeling of various queueing scenarios while bridging quantum computing and classical queueing theory. Notably, our quantum method shows potential advantages over classical simulations, particularly in high-traffic scenarios. This quantum simulation approach opens new possibilities for analyzing complex queueing systems, potentially outperforming classical methods in challenging scenarios and paving the way for quantum-enhanced operations research. The method was implemented and tested across low-, moderate-, and high-traffic scenarios, comparing quantum simulations with both theoretical formulas and classical simulations. Results demonstrate high agreement between quantum computations and theoretical predictions, with relative errors below 0.002 for effective arrival rates in high-traffic scenarios. As the number of qubits increases, we observe rapid convergence to theoretical values, with relative errors decreasing by up to two orders of magnitude in some cases. Sensitivity analysis reveals optimal parameter regions yielding errors lower than 0.001.

**Keywords.** *amplitude amplification*, *Grover diffusion operator*, *stochastic processes, Markovian queue system.*

## 1. INTRODUCTION

Quantum computing (QC) represents a revolutionary frontier in computational science, leveraging quantum mechanics principles to process information in fundamentally different ways from classical computers (Gyongyosi & Imre, 2019). QC exploits phenomena such as superposition and entanglement, enabling the simultaneous exploration of multiple solution paths and potentially solving problems intractable for classical computers (Ying, 2010). This capability promises to have transformative applications across finance, healthcare, and artificial intelligence (Piattini et al., 2021).

Queueing theory, a branch of operations research, provides a mathematical framework for studying waiting lines, or queues (Sundarapandian, 2009). Classical queueing systems typically involve servers attending to customers arriving via a stochastic process. These systems are characterized by factors such as arrival process, service time distribution, number of servers, queue capacity, and service discipline. The M/M/n queue model, in which arrival times are Poisson distributed and service times are exponentially distributed, forms a foundation for many queueing analyses (Ibe, 2013).

As quantum computing advances, researchers have begun exploring quantum analogues of classical queueing systems, combining principles from quantum mechanics, information theory, and classical queueing theory. This convergence offers potential for novel insights and more efficient methods of queue management and optimization in quantum computing environments.

This study presents a novel quantum approach for simulating single-server Markovian queues, addressing key challenges in quantum computing and queueing theory. Our work makes several important contributions: (1) We develop a method to simulate M/M/1 queues using quantum circuits, making quantum computing more accessible to researchers in operations research and queueing theory. (2) We introduce a dynamic amplification technique that adapts to queue traffic, potentially improving simulation efficiency. (3) We design custom-parameterized quantum gates for arrival and service processes, providing a flexible framework for modeling various queueing scenarios. Through these innovations, we aim to bridge the gap between classical queueing theory and quantum computing, potentially enabling more efficient analysis of complex queueing systems.

The remainder of this paper is structured as follows: Section 2 reviews recent advances in quantum computing and queueing theory. Section 3 details our dynamic amplification method, along with its implementation and proof of correctness. Section 4 describes our empirical study, including experimental procedures and an illustrative example. Section 5 presents results, with a comparison of the results of our quantum simulation to theoretical M/M/1 formulas across different traffic scenarios. Finally, Section 6 discusses the implications of our findings and suggests future research directions.

## 2. RELATED WORK

As the field advances, quantum algorithms are being developed to harness quantum systems' power. Although no fundamental functional distinction currently exists between classical and quantum computers, quantum gates have enabled the transformation of classical machine learning algorithms into quantum counterparts (Benedetti et al., 2019), giving rise to quantum machine learning (QML) (Buffoni & Caruso, 2021; Peretz & Koren, 2024; Alchieri et al., 2021).

Quantum approximation algorithms (QAAs) have emerged as a promising approach to complex computational problems (Koren & Peretz, 2024; Koren et al., 2023), showing significant improvements in solving optimization issues such as integer factorization (Hadfield et al., 2019) and combinatorial optimization (Zhang et al., 2022).The ongoing development of quantum computing and its algorithms is expected to drive innovations in cryptography, machine learning, and scientific simulations, potentially catalyzing significant progress in science and technology (Choi & Kin, 2019; Willsch et al., 2020; Koren & Peretz, 2024).

The Grover search algorithm is designed for searching unstructured data (Cui & Fan, 2010). In contrast to a classical computer search algorithm, the Grover search algorithm is based only on square root operations for searching (Mandal et al., 2014). This algorithm is built upon the Grover diffusion operator, which flips the amplitudes of all states around their mean (Shi et al., 2017). This process, known as "inversion around the mean," amplifies the marked state's amplitude while decreasing the unmarked states' amplitudes (Tulsi, 2015). Through repeated application of this operator, the algorithm progressively concentrates the amplitude on the target state, thereby significantly increasing the probability of identifying it (Jang et al., 2021).

Queueing theory, a branch of operations research, provides a mathematical framework for studying waiting lines, or queues (Sundarapandian, 2009; Shortle et al., 2018). Classical queueing systems typically involve servers attending to customers arriving via a stochastic process. These systems are characterized by factors such as arrival process, service time distribution, number of servers, queue capacity, and service discipline (Adan & Resing, 2002). When it is true both that the arrival times are Poisson distributed and that the service times are exponentially distributed, the system can be modeled as a Markov chain, a foundation for many queueing analyses (Armero, 1994; Ibe, 2013).

Stochastic processes, particularly in queueing theory, present significant computational challenges due to inherent randomness in arrival and service times (Montanaro, 2016). These challenges often result in complex optimization problems that classical computing methods struggle to solve efficiently (Alaghi et al., 2017; Nachman et al., 2020).

As quantum computing advances, researchers have begun exploring quantum analogues of classical queueing systems, combining principles from quantum mechanics, information theory (Horodecki et al., 2022), and classical queueing theory. Key developments include random quantum allocation for analyzing waiting-time distributions in multiple-server Markovian processor sharing queues (Braband & Schaßberger, 1993), discrete-time quantum Markov chain models for quantum queues (Gawron et al., 2013; Guimarães et al., 2023), and quantum walk–based queue models for optimizing throughput in quantum queueing networks (Kurzyk and Gawron, 2014).

Recent research has explored Markovian quantum systems in various contexts (Mikki, 2023; Bougron et al., 2022), contributing to our understanding of quantum state evolution and entropy statistics in queueing-like scenarios (Zurek, 2022). Additionally, techniques for quantum state estimation, such as the quantum Bayesian approach and hybrid quantum-classical methods, have been developed, with potential applications in quantum queueing systems (Yang et al., 2018; Madsen et al., 2021).

Quantum computers leverage qubits that can exist in multiple states simultaneously due to superposition (Nielsen & Chuang, 2010). This unique capability allows quantum systems to process numerous potential outcomes concurrently, making them well-suited for problems involving probabilistic outcomes. In queueing theory, this quantum advantage can be harnessed to simultaneously evaluate different queue configurations or service protocols, potentially optimizing system performance more quickly than classical methods (Prudencio & Fischer, 2019).

The application of quantum computing to stochastic process optimization offers several promising avenues (Hassija et al., 2020), including more effective evaluation of probabilistic outcomes (Milz & Modi, 2021; Korzekwa & Lostaglio, 2021), significant speed-ups in semidefinite programming, and improved operational efficiency and strategic planning (Brandao et al., 2017; Dahlsten et al., 2020).

# 3. QUANTUM-ENHANCED ADAPTIVE QUEUEING SIMULATION FRAMEWORK

This section presents and describes a novel quantum approach for simulating single-server Markovian queues using dynamic amplification. Our dynamic amplification method leverages quantum superposition and interference to simulate M/M/1 queues more efficiently than classical methods. The key innovation lies in adapting the quantum state amplification based on the current queue traffic using the quantum amplitude amplification technique, a generalization of Grover's algorithm. We begin by introducing the custom-parameterized quantum gates designed for this study. Next, we detail the quantum method, including its logic and circuit implementation. Finally, we provide a comprehensive analysis of the method's implementation, demonstrate its correctness, and provide an upper bound for the circuit error. Table I summarizes the notation used throughout this study.

**Table I**. The notation used in this study.

| Symbol | Remarks |
|---|---|
| $\lambda$ | Arrival rate |
| $\lambda_{\text{eff}}$ | Effective arrival rate |
| $\mu$ | Service rate |
| $k$ | Current queue length |
| $\Delta t$ | Time interval |
| $M$ | A set of quantum states |
| $T$ | Number of steps in the quantum circuit |
| $\alpha, \beta \in \mathbb{R}^+$ | Amplification parameters for arrival and service gates, respectively |
| $\varepsilon_0, \varepsilon_1$ | Amplification interval such that $0 < \varepsilon_0 < \varepsilon_1 < 1$ |

## 3.1. Parameterized quantum gates

This study defines new parameterized quantum gates for the arrival and service processes in an M/M/1 queue simulation. Let $A(\lambda, k, \alpha, \Delta t)$ and $S(\mu, k, \alpha, \Delta t)$ be the quantum arrival and service gates, respectively. These gates are based on the rotation around the $Y$ axis gate:

$$RY(\theta) = e^{-i\frac{\theta}{2}Y} = \begin{bmatrix} \cos\left(\frac{\theta}{2}\right) & -\sin\left(\frac{\theta}{2}\right) \\ \sin\left(\frac{\theta}{2}\right) & \cos\left(\frac{\theta}{2}\right) \end{bmatrix}$$

where the angles for each gate are:
$$\theta_A = 2\arcsin\left(\sqrt{\min(\lambda(1+\alpha k)\Delta t, 1)}\right)$$
$$\theta_S = 2\arcsin\left(\sqrt{\min(\mu(1+\beta k)\Delta t, 1)}\right)$$

Our quantum gates $A(\lambda, k, \alpha, \Delta t)$ and $S(\mu, k, \alpha, \Delta t)$ are designed to capture M/M/1 queue dynamics with quantum amplification. The inclusion of $k$ allows adaptation to current queue length, while the amplification parameters $\alpha$ and $\beta$ enhance sensitivity to queue changes, potentially accelerating convergence to steady state. These parameters are crucial and can be tuned based on queue characteristics.

A new operation that is defined on a quantum circuit is required to be unitary[1], since any physical operation on a state using to evolve it (Bennett et al., 1997). Since $A(\lambda, k, \alpha, \Delta t)$ and $S(\mu, k, \alpha, \Delta t)$ are rotation matrices, they are inherently unitary. For full proof of concept, see Section 3.3.

### 3.2. Quantum logic and circuit

The method starts with quantum state initialization and creates a quantum circuit with $n = [\log(k)] + 1$ qubits, where $n$ is chosen to represent up to $2^n - 1$ customers in the queue. The circuit is initialized to the state $|0\rangle^{\otimes n}$, representing an empty queue. The procedure consists of two main components:

1. **Grover operator for marked states:** The Grover operator is constructed to amplify the probability of observing certain marked states. These marked states are dynamically determined based on the queue utilization, $\rho = \frac{\lambda}{\mu} < 1$, and two threshold parameters, $\varepsilon_0$ and $\varepsilon_1$ (where $0 < \varepsilon_0 < \varepsilon_1 < 1$). This dynamic selection allows the quantum circuit to adapt to different queue conditions, supporting a more realistic simulation of the M/M/1 queue system. The function selects the marked states as follows: (1) For low utilization ($\rho < \varepsilon_0$), it amplifies states representing an empty or near-empty queue. (2) For moderate utilization ($\varepsilon_0 \leq \rho < \varepsilon_1$), it amplifies states representing a queue with a small number of customers. (3) For high utilization ($\rho \geq \varepsilon_1$), it amplifies states representing a queue with more customers.

    Once the marked states are determined, the custom Grover operator $G'$ is constructed as:
    $$G = (2|\psi\rangle\langle\psi| - I)(2|m\rangle\langle m| - I)$$
    where $|\psi\rangle = \frac{1}{\sqrt{2^n}} \sum_{i=0}^{2^n-1} |i\rangle$ is the equal superposition state and $|m\rangle = \frac{1}{\sqrt{|M|}} \sum_{s \in M} |s\rangle$ is the superposition of marked states.

2. **Quantum M/M/1 simulation with dynamic amplification:** At first, the method initialized a quantum circuit of $n = [\log(k)] + 1$ qubits, where $k$ represents the maximum queue capacity. The initial state was set to 0, represented as $|\psi\rangle \leftarrow |0\rangle^{\otimes n}$. Then, the method determined the marked states ($M$), basing its selection on the queue parameters $\lambda$ and $\mu$ and the expected traffic. Using these marked states, the method created the Grover operator $G$ to amplify these states.

    The simulation then proceeded through $T$ time steps. Let $A'$ be the arrival gate, defined as $A' = \bigotimes_{i=0}^{n-1} A(\lambda, s_t, \alpha, \Delta t)$, and $S'$ be the service gate, defined as $S' = \bigotimes_{i=0}^{n-1} S(\mu, s_t, \beta, \Delta t)$, where $s_t$ represents the current simulation step. For each step from 0 to $T - 1$, the method applied these gates to the quantum state, and

---

[1] A matrix $U$ is unitary if and only if its conjugate transpose equal to its inverse, i.e., $U^\dagger = U^{-1}$.

the resulting state vector was $|\psi\rangle \leftarrow S' \cdot A'|\psi\rangle$. Following this, the simulation applied $G$ total of $K$ times to amplify the desired states.

The arrival and service gates were dynamically adjusted based on the current step of the simulation, allowing for time-dependent rates controlled by parameters $\alpha$ and $\beta$. This dynamic adjustment implemented the core of the amplification strategy. After completing all $T$ steps, the method measured the final quantum state $|\psi\rangle$, providing a representation of the queue's behavior over the simulated period.

Figure 1 provides a schematic representation of our quantum circuit for M/M/1 queue simulation. The circuit illustrates the initialization of qubits, the application of parameterized gates $A$ and $S$, and the implementation of the Grover operator for state amplification.

**DetermineMarkedStates($\lambda, \mu, n, \varepsilon_0, \varepsilon_1$)**
- $\rho \leftarrow \frac{\lambda}{\mu}$
- $M \leftarrow \emptyset$
- If $\rho < \varepsilon_0$
  - **FOR** $S_i = 0$ to $\left\lceil \frac{2^n - 1}{3} \right\rceil$ **DO** $M \leftarrow M \cup \{S_i\}$
- else if $\varepsilon_0 \leq \rho < \varepsilon_1$
  - **FOR** $S_j = \left\lceil \frac{2^n - 1}{3} \right\rceil$ to $\left\lceil \frac{2(2^n - 1)}{3} \right\rceil$ **DO** $M \leftarrow M \cup \{S_j\}$
- Else
  - **FOR** $S_k = \left\lceil \frac{2(2^n - 1)}{3} \right\rceil$ to $(2^n - 1)$ **DO** $M \leftarrow M \cup \{S_k\}$
- Return $M$

**CreateGroverOperator(n, M)**
- $|m\rangle = \frac{1}{\sqrt{|M|}} \sum_{s \in M} |s\rangle$
- $R_m \leftarrow (2|m\rangle\langle m| - I)$
- $|\psi\rangle = \frac{1}{\sqrt{2^n}} \sum_{i=0}^{2^n - 1} |i\rangle$
- $R_\psi \leftarrow (2|\psi\rangle\langle\psi| - I)$
- Return $R_\psi \cdot R_m$

**QuantumMM1Simulation($\lambda, \mu, \alpha, \beta, \Delta t, T, n, \varepsilon_0, \varepsilon_1$)**
- Circuit ← QuantumCircuit($n, n$)
- $|\psi\rangle \leftarrow |0\rangle^{\otimes n}$
- $M \leftarrow$ **DetermineMarkedStates($\lambda, \mu, n, \varepsilon_0, \varepsilon_1$)**
- $G \leftarrow$ **CreateGroverOperator($n, M$)**
- For $s\_t = 0$ to $T - 1$
  - $A' \leftarrow \otimes_{i=0}^{n-1} A(\lambda, s_t, \alpha, \Delta t)$
  - $S' \leftarrow \otimes_{i=0}^{n-1} S(\mu, s_t, \beta, \Delta t)$
  - $|\psi\rangle \leftarrow S' \cdot A' \cdot |\psi\rangle$
  - For $j = 0$ to $k - 1$
    - $|\psi\rangle \leftarrow G|\psi\rangle$
- Measure $|\psi\rangle$

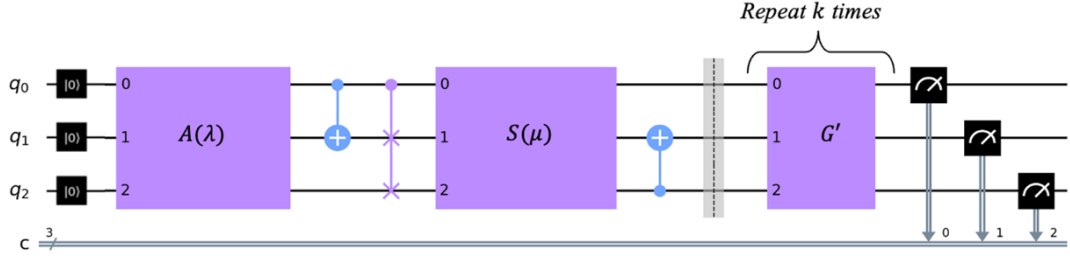

**Figure 1.** The proposed quantum circuit

### 3.3. Correctness

*Theorem 1.* The arrival and service gates, $A(\lambda, k, \alpha, \Delta t)$ and $S(\mu, k, \alpha, \Delta t)$, are unitary operations.

*Proof.* Let $A(\lambda, k, \alpha, \Delta t)$ be the quantum gate representing arrival event in M/M/1 queue system. By the arrival gate definition:

$$A = RY(\theta) = \begin{bmatrix} \cos\left(\frac{\theta}{2}\right) & -\sin\left(\frac{\theta}{2}\right) \\ \sin\left(\frac{\theta}{2}\right) & \cos\left(\frac{\theta}{2}\right) \end{bmatrix}$$

In quantum interpretation, the effect of the $RY(\theta)$ gate on a qubit in state $|0\rangle$ is:

$$RY(\theta)|0\rangle = \cos\left(\frac{\theta}{2}\right)|0\rangle + \sin\left(\frac{\theta}{2}\right)|1\rangle$$

The probability of measuring $|1\rangle$ (representing an arrival) is:

$$P(|1\rangle) = \sin^2\left(\frac{\theta}{2}\right) = p'$$

$$\frac{\theta}{2} = \arcsin\left(\sqrt{p'}\right)$$

$$\theta = 2\arcsin\left(\sqrt{p'}\right)$$

For simplification, let $p' = \min(\lambda(1 + \alpha k)\Delta t, 1)$; thus:

$$\cos\left(\frac{2\arcsin\left(\sqrt{\min(\lambda(1 + \alpha k)\Delta t, 1)}\right)}{2}\right) = \sqrt{1 - \min(\lambda(1 + \alpha k)\Delta t, 1)}$$

$$\sin\left(\frac{2\arcsin\left(\sqrt{\min(\lambda(1 + \alpha k)\Delta t, 1)}\right)}{2}\right) = \sqrt{\min(\lambda(1 + \alpha k)\Delta t, 1)}$$

and the obtained arrival gate is:

$$A = \begin{bmatrix} \sqrt{1-p'} & -\sqrt{p'} \\ \sqrt{p'} & \sqrt{1-p'} \end{bmatrix}$$

This gate is unitary (i.e., $A^\dagger = A^{-1}$):

$$A^\dagger = \begin{bmatrix} \sqrt{1-p'} & \sqrt{p'} \\ -\sqrt{p'} & \sqrt{1-p'} \end{bmatrix}$$

$$A^{-1} = \frac{1}{\det(A)} \cdot \operatorname{adj}(A) = \frac{1}{\begin{vmatrix} \sqrt{1-p'} & -\sqrt{p'} \\ \sqrt{p'} & \sqrt{1-p'} \end{vmatrix}} \cdot \begin{pmatrix} \sqrt{1-p'} & \sqrt{p'} \\ -\sqrt{p'} & \sqrt{1-p'} \end{pmatrix} =$$

$$= \frac{1}{(1-p')-(-p')} \cdot \begin{pmatrix} \sqrt{1-p'} & \sqrt{p'} \\ -\sqrt{p'} & \sqrt{1-p'} \end{pmatrix} = \begin{pmatrix} \sqrt{1-p'} & \sqrt{p'} \\ -\sqrt{p'} & \sqrt{1-p'} \end{pmatrix} = A^\dagger$$

Therefore, $A^\dagger = A^{-1}$ and $A^\dagger A = I$, proving that $A$ is unitary.

Note: The proof for the service gate $S$ would follow a similar structure, with $\lambda$ replaced by $\mu$ and $\alpha$ by $\beta$.

*Theorem 2.* The arrival gate $A(\lambda, k, \alpha, \Delta t)$ models a Poisson process in a Markovian queue with dynamic rate $\lambda(1 + \alpha k)$.

*Proof.* Let $N(t)$ be a Poisson process with rate $\lambda(1 + \alpha k)$ such that the probability of $n$ arrivals in time $t$ can be expressed as:

$$P(N(t) = n) = e^{-\lambda(1+\alpha k)t} \cdot \frac{(\lambda(1+\alpha k)t)^n}{n!}$$

First, note that when $\alpha \to 0$ and $k$ is finite:

$$\lim_{\alpha \to 0} \lambda(1 + \alpha k) = \lambda$$

This limit demonstrates that our model converges to the standard M/M/1 queue when there is no amplification, ensuring consistency with the classical model. To show that the arrival process follows a Poisson process with rate $\lambda(1 + \alpha k)$, we need to demonstrate three key properties:

1. The probability of exactly one arrival in a small time interval.

From Theorem 1, the probability of an arrival in a small time interval $\Delta t$ is:

$$P(N(\Delta t = 1)) = \sin^2\left(\frac{\theta_A}{2}\right) = \min(\lambda(1+\alpha k)\Delta t, 1)$$

Thus, for small $\Delta t$, this probability is approximately $\lambda(1 + \alpha k)\Delta t$. Note that when $\lambda(1 + \alpha k)\Delta t$ approaches or exceeds 1, we use the minimum function to cap the probability, ensuring that our model remains physically meaningful.

Consider a small interval $[t, t + \Delta t]$. The probability of there being one arrival in this interval is $\lambda(1 + \alpha k)\Delta t + o(\Delta t)$, and the probability of there being more than one arrival is $o(\Delta t)$. This follows from the properties of the quantum gate, as the probability of multiple state changes in a small interval is negligible.

2. The independence of arrivals in non-overlapping intervals.

By the definition of quantum circuit, the quantum gate $A$ operates independently for each application, ensuring that arrivals in non-overlapping intervals are independent events.

3. <u>The exponential distribution of interarrival times.</u>

Let $T$ be the time until the next arrival. By properties of Poisson process, $T$ is an exponential distributed random variable, i.e., we need to show that:
$$P(T > t) = e^{-\lambda(1+\alpha k)t}$$
For a small interval $\Delta t$, the probability of there being no arrival is:
$$P(\text{no arrival in } \Delta t) = 1 - \lambda(1 + \alpha k)\Delta t + o(\Delta t)$$
The probability of no arrival in time $t$ can be expressed as:
$$P(\text{no arrival in } \Delta t) = \lim_{n \to \infty} \left(1 - \frac{\lambda(1 + \alpha k)t}{n}\right)^n$$
Using the definition of $e^x$, we get:
$$e^x = \lim_{n \to \infty} \left(1 + \frac{x}{n}\right)^n$$
$$\lim_{n \to \infty} \left(1 - \frac{\lambda(1 + \alpha k)t}{n}\right)^n = e^{-\lambda(1+\alpha k)t}$$
This is the cumulative distribution function of an exponential distribution with rate $\lambda(1 + \alpha k)$. We have demonstrated that the arrival process modeled by gate $A$ satisfies the key properties of a Poisson process. Furthermore, the rate of this Poisson process is dynamically adjusted based on the current queue length, given by $\lambda(1 + \alpha k)$. This proves that the arrival gate $A$ accurately models a Poisson process with dynamic rate in a Markovian queue.

*Theorem 3.* The service gate $S(\mu, k, \alpha, \Delta t)$ models exponential service times in a Markovian queue with dynamic rate $\mu(1 + \beta k)$.

*Proof.* Let $Y \sim \exp(\mu)$ be the distribution of service time. For the service times to follow an exponential distribution with rate $\mu(1 + \beta k)$, we need to show that:
$$P(Y > t) = e^{-\mu(1+\beta k)t}$$
where $\mu(1 + \beta k)$ is the dynamically adjusted service rate. From Theorem 1, the probability of a service completion in a small-time interval $\Delta t$ is
$$P(\text{service completion in } \Delta t) = \sin^2\left(\frac{\theta_S}{2}\right) = \min(\mu(1 + \beta k)\Delta t, 1)$$
For small $\Delta t$, this probability is approximately $\mu(1 + \beta k)\Delta t$. The probability that the service time exceeds $t$ units is the probability of no service completion occurring in time $t$:
$$P(Y > t) = \lim_{\Delta t \to 0} (1 - \mu(1 + \beta k)\Delta t)^{\frac{t}{\Delta t}}$$
Taking the natural logarithm of both sides:
$$\ln P(Y > t) = \ln\left(\lim_{\Delta t \to 0} (1 - \mu(1 + \beta k)\Delta t)^{\frac{t}{\Delta t}}\right)$$
$$\ln P(Y > t) = \lim_{\Delta t \to 0} \left(\ln\left((1 - \mu(1 + \beta k)\Delta t)^{\frac{t}{\Delta t}}\right)\right)$$
$$\ln P(Y > t) = \lim_{\Delta t \to 0} \left(\frac{t}{\Delta t}\ln\left((1 - \mu(1 + \beta k)\Delta t)\right)\right)$$

Applying L'Hopital's rule:
$$\ln P(Y > t) = \lim_{\Delta t \to 0} \left( t \cdot \frac{-\mu(1+\beta k)}{1 - \mu(1+\beta k)\Delta t} \right)$$

We can rewrite this as:
$$\ln P(Y > t) = -\mu(1+\beta k)t \cdot \lim_{\Delta t \to 0} (1 - \mu(1+\beta k)\Delta t)^{-1}$$

Given that $e^x = \lim_{n \to \infty} \left(1 + \frac{x}{n}\right)^n$, and $n = \frac{1}{\Delta t}$ and $x = -\mu(1+\beta k)$ satisfies $\Delta t \to 0, n \to \infty$, thus:
$$\frac{x}{n} = -\mu(1+\beta k)\Delta t$$

Assign it in the limit:
$$\lim_{\Delta t \to 0} (1 - \mu(1+\beta k)\Delta t)^{-1} = e^{\mu(1+\beta k) \cdot 0} = e^0 = 1$$
$$\ln P(Y > t) = -\mu(1+\beta k)t$$
$$P(Y > t) = e^{-\mu(1+\beta k)t}$$

This is the service function of an exponential distribution with rate $\mu(1+\beta k)$. The queue length affects the service process by dynamically adjusting the service rate. As $k$ increases, the effective service rate $\mu(1+\beta k)$ increases, modeling a system in which higher queue occupancy leads to faster service.

Note that when $\mu(1+\beta k)\Delta t$ approaches or exceeds 1, we use the minimum function to cap the probability, ensuring that our model remains physically meaningful.

### 3.4. Error bound for quantum effective arrival rate simulation

Let $\lambda_{\text{eff}}^C$ be the classical effective arrival rate of an M/M/1/K queue, defined as:
$$\lambda_{\text{eff}}^C = \lambda(1 - P_K)$$

where $P_K$ is the probability of the system being in state $K$ (i.e., full capacity). For an M/M/1/K queue, it follows that:
$$P_K = \frac{\rho^K (1-\rho)}{1 - \rho^{K+1}}$$

Substituting $K = 2^n - 1$, we get:
$$\lambda_{\text{eff}}^C = \lambda \left(1 - \frac{\rho^{2^n - 1}(1-\rho)}{1 - \rho^{2^n}}\right)$$

The quantum simulation approximates the state probabilities using amplitude amplification. Let $|\psi\rangle$ be the final quantum state after $T$ time steps and $K$ Grover iterations. The probability of measuring state $|K\rangle$ is $|\langle K|\psi\rangle|^2$. The quantum effective arrival rate can be expressed as:
$$\lambda_{\text{eff}}^Q = \lambda(1 - |\langle 2^n - 1|\psi\rangle|^2)$$

The error, denoted $e$, is the absolute difference between the quantum and the classical effective arrival rate:
$$e = |\lambda_{\text{eff}}^C - \lambda_{\text{eff}}^Q| = \lambda \left(\frac{\rho^{2^n - 1}(1-\rho)}{1 - \rho^{2^n}} - |\langle 2^n - 1|\psi\rangle|^2\right)$$

For a quantum state prepared with Grover iterations, the error in estimating the probability of a marked state is $O\left(\frac{1}{\sqrt{2^n}}\right)$ (Brassard et al., 2002); thus:

$$e = |\lambda_{\text{eff}}^C - \lambda_{\text{eff}}^Q| \leq \frac{\lambda \rho^{2^n}}{1 - \rho^{2^n}} + O\left(\frac{1}{\sqrt{2^n}}\right)$$

For a stable queue, $\rho < 1$, it follows that:

$$\lim_{n \to \infty} \frac{\lambda \rho^{2^n}}{1 - \rho^{2^n}} = \frac{\lambda \cdot 0}{1 - 0} = 0$$

Thus,

$$e \approx O\left(\frac{1}{\sqrt{2^n}}\right)$$

Note that in case of $n \to \infty$, the error reaches 0. However, the $O\left(\frac{1}{\sqrt{2^n}}\right)$ term decreases more slowly than the first term, so it becomes the dominant source of error for large $n$.

## 4. EMPIRICAL STUDY
### 4.1. Experimental procedure
In the experiments, the predefined parameters were set as equal for all scenarios. The following describes the setup for each experiment:

1. Scenarios: Three traffic scenarios were analyzed using the quantum simulation, classical simulation, and theoretical M/M/1/K formulas:
    - Low traffic: $\lambda = 1, \mu = 10$ ($\rho = 0.1$)
    - Medium traffic: $\lambda = 5, \mu = 10$ ($\rho = 0.5$)
    - High traffic: $\lambda = 9.5, \mu = 10$ ($\rho = 0.95$)

    For each scenario, we compared the result of the quantum method with the theoretical values and with the result obtained using the discrete event simulator (DES) for classical computation (Babulak & Wang, 2010).

2. Quantum circuits were initialized with $n = \{2,3,4,5,6,7,8,9\}$ qubits, representing a system capacity of $2^n - 1$. The initial state for the quantum system is set to $|0\rangle$, representing an empty queue.

3. We calculated and compared the following metrics:
    - Total number of customers in the system: $E[L_s] = \frac{\rho}{1-\rho} - \frac{2^n \cdot \rho^{2^n}}{1-\rho^{2^n}}$.
    - Total number of customers waiting in the queue: $[L_q] = E[L_s] - (1 - P_0)$.
    - Effective arrival rate: $\lambda_{\text{eff}} = \lambda(1 - P_{(2^n-1)})$
    - The expected time a customer stays in the system: $E[W_s] = \frac{E[L_s]}{\lambda_{\text{eff}}}$
    - The expected time a customer stays in the waiting queue: $E[W_q] = \frac{E[L_s]}{\lambda_{\text{eff}}} - \frac{1}{\mu}$.

4. The relative error between quantum and theoretical results was computed for each metric.

5. For all scenarios, amplification parameters $\alpha = \beta = 0.1$ were used initially. The number of simulation steps was set to 1000 for each run. To examine the amplification parameters, we conducted a comprehensive sensitivity analysis to evaluate the robustness of our quantum simulation method across three traffic scenarios: low, moderate, and high. The analysis focused on the impact of varying the amplification parameters $\alpha$ and $\beta$ on the effective arrival rate. We used the following setup:

- Number of qubits: six (allowing for up to 63 customers in the queue)
- Time step ($\Delta t$): 0.1
- Number of simulation steps: 1000
- Number of shots: 10,000 (for statistical significance)

We independently varied $\alpha$ and $\beta$ from 0.01 to 0.15 in increments of 0.01. For each combination, we ran the quantum simulation and calculated the relative error between the quantum and theoretical effective arrival rate. This process was repeated for all three traffic scenarios.

**4.2. Simple demonstration**

Let $\lambda = 2$ be the arrival rate and let $\mu = 3$ be the service rate of a single-server Markovian queue system. This demonstration uses $n = 3$ qubits (i.e., allowing for up to 7 customers in the queue); thus, the system can be defined as M/M/1/7. For simplification, we used $\alpha = 0.1, \beta = 0.1$ as amplification parameters and $\varepsilon_0 = 0.3, \varepsilon_1 = 0.7$ for utilization thresholds. The utilization $\rho = \frac{\lambda}{\mu} = \frac{2}{3} \approx 0.67$, which falls in the moderate utilization range ($\varepsilon_0 \leq \rho < \varepsilon_1$).

Calculating the expected value of customers in the system using the theoretical formula and classical computation:

$$E[L_s] = \frac{\rho}{1-\rho} - \frac{2^n \cdot \rho^{2^n}}{1-\rho^{2^n}} = \frac{\frac{2}{3}}{\frac{1}{3}} - \frac{8 \cdot \left(\frac{2}{3}\right)^8}{1 - \left(\frac{2}{3}\right)^8} = 2 - \frac{2048}{6305} = 1.675$$

The quantum simulation initializes the quantum state as $|\psi\rangle = |000\rangle$, representing an empty queue. First, we determine the marked states:

$$M = \{|001\rangle, |010\rangle, |011\rangle\}$$

Next, the procedure creates the Grover operator:

$$|m\rangle = \frac{1}{\sqrt{3}}(|001\rangle + |010\rangle + |011\rangle)$$

$$G = (2|\psi\rangle\langle\psi| - I)(2|m\rangle\langle m| - I)$$

For the simulation, we apply the arrival and service operators (defined in Section 3.1) with the following angles:

$$\theta_A(k) = 2\arcsin\left(\sqrt{\min(2(1+0.1k) \cdot 0.1, 1)}\right)$$

$$\theta_S(k) = 2\arcsin\left(\sqrt{\min(3(1+0.1k) \cdot 0.1, 1)}\right)$$

When the queue is empty, i.e., $k = 0$, we get:

$$\theta_A(0) = 2\arcsin\left(\sqrt{\min(0.2, 1)}\right) = 2\arcsin(\sqrt{0.2}) = 0.927$$

$$\theta_S(0) = 2\arcsin\left(\sqrt{\min(0.3, 1)}\right) = 2\arcsin(\sqrt{0.3}) = 1.159$$

Note that the calculations are in radians. Calculating the state after the first time step yields:

$$A(2,0,0.1,0.1) = \begin{bmatrix} 0.894 & -0.447 \\ 0.447 & 0.894 \end{bmatrix}$$

$$S(3,0,0.1,0.1) = \begin{bmatrix} 0.836 & -0.547 \\ 0.547 & 0.836 \end{bmatrix}$$

$$S(\mu,0,\beta,\Delta t) \cdot A(\lambda,0,\alpha,\Delta t) = \begin{bmatrix} 0.502 & -0.862 \\ 0.862 & 0.502 \end{bmatrix}$$

Given an empty queue:

$$|\psi_1\rangle = \begin{bmatrix} 0.502 & -0.862 \\ 0.862 & 0.502 \end{bmatrix} \begin{bmatrix} 1 \\ 0 \end{bmatrix} = \begin{bmatrix} 0.502 \\ 0.862 \end{bmatrix} = 0.502|000\rangle + 0.862|001\rangle$$

After one time step, there is an approximately 25.2% ($0.502^2 \approx 0.252$) probability of the queue being empty and an approximately 74.8% ($0.862^2 \approx 0.748$) probability of the queue having one customer. After applying the operators 100 times, we measure the final state:

$$|\psi_{10}\rangle = \sqrt{0.201}|000\rangle + \sqrt{0.280}|001\rangle + \sqrt{0.313}|010\rangle + \sqrt{0.109}|011\rangle + \sqrt{0.097}|100\rangle$$

This final state represents the probability distribution of different queue lengths after the simulation. To analyze the results, we can calculate the expected queue length:

$$Q(E[L_s]) = 0 \cdot 0.201 + 1 \cdot 0.280 + 2 \cdot 0.313 + 3 \cdot 0.109 + 4 \cdot 0.097 = 1.621$$

The recorded difference between the classical and quantum covariance is 0.054. Figure 2 presents the final state evolution of $|\psi\rangle$ over three qubits in Bloch sphere representation, with $|\psi_i\rangle$ being the state-vector in the $i^{\text{th}}$ time step, where $0 \leq i \leq 3$.

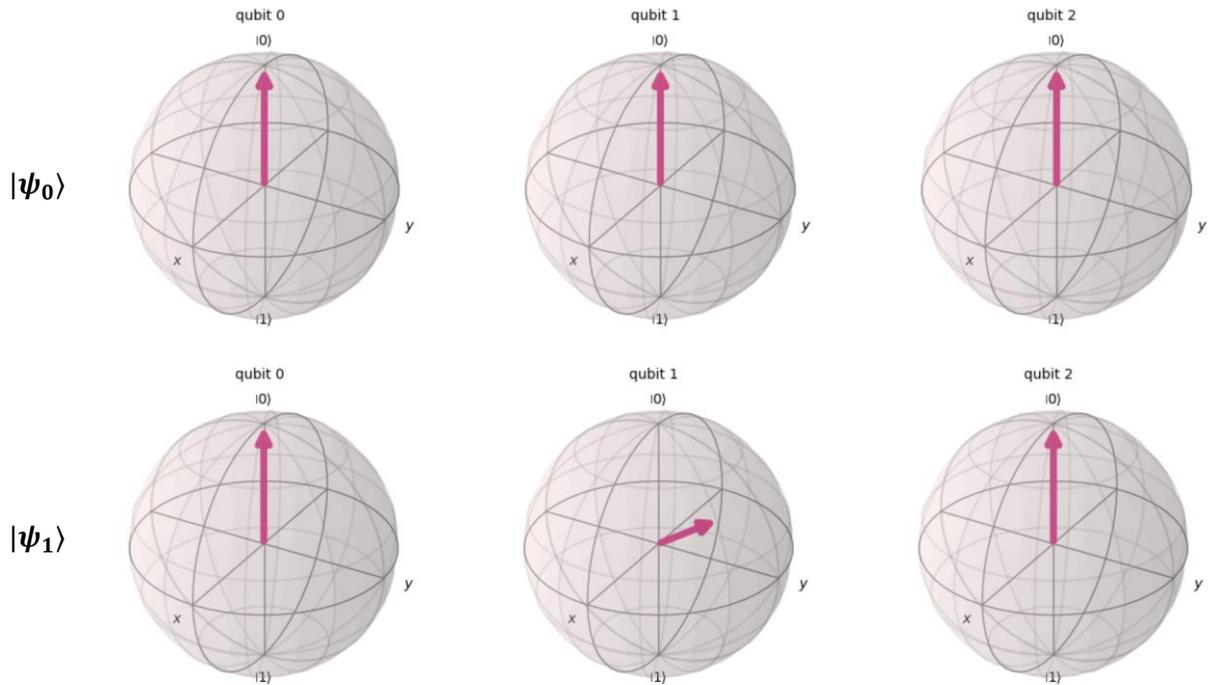

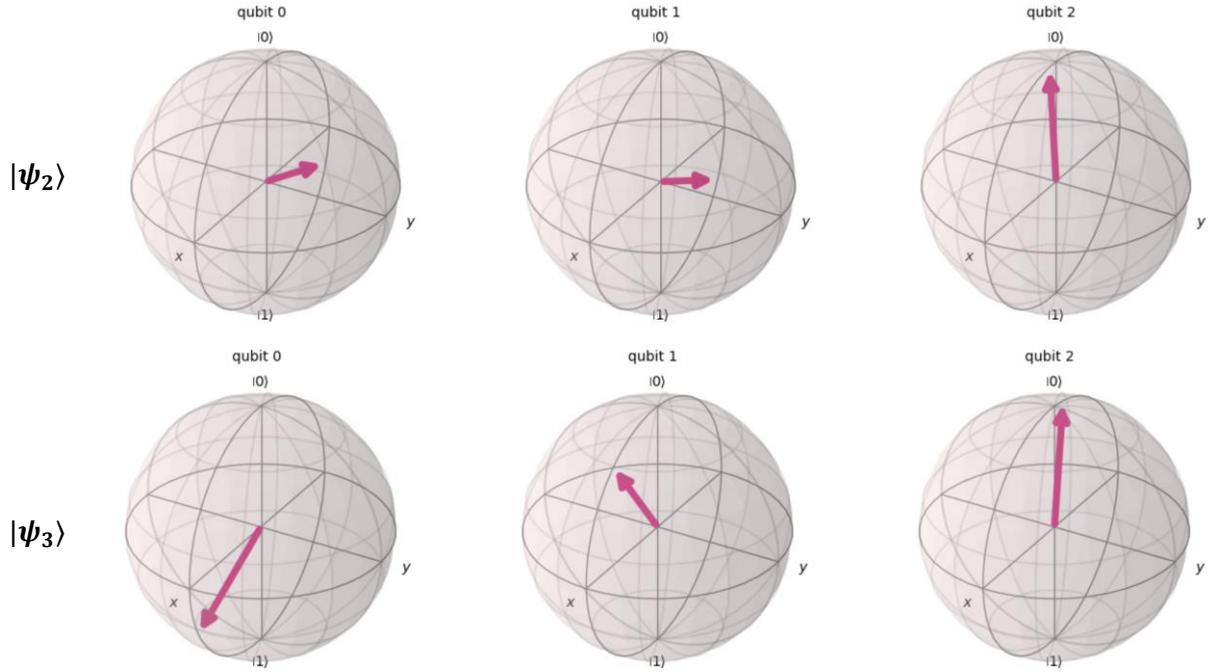

**Figure 2.** Evolution of quantum state in a three-qubit M/M/1 queue simulation

## 5. RESULTS

This section presents the outcomes of our quantum simulation for single-server Markovian queues and compares them with theoretical predictions. We evaluate the performance of our dynamic amplification method across three traffic scenarios and compare queues measures as defined in Section 4.1. First, we provide a detailed comparison between our quantum simulation results and classical theoretical calculations for these metrics. Then, we present a sensitivity analysis to assess the robustness of our method under varying parameters.

### 5.1. A comparison between quantum, classical, and theoretical approaches

For this analysis, we used a quantum circuit with five qubits, corresponding to a maximum system capacity of $K = 31$ customers (a broader analysis exploring the impact of number of qubits is presented later in this section). Table II summarizes the quantum, theoretical, and DES results for each metric across the three scenarios. Following this, we present a more detailed analysis of how $\lambda_{\text{eff}}$ varies with the number of qubits, providing insights into the convergence and accuracy of our quantum simulation method under different traffic conditions.

|  | Measure | Quantum | Theoretical | DES |
|---|---|---|---|---|
| **Low Traffic** | $L_q$ | 0.097 | 0.011 | 0.011 |
|  | $L_s$ | 0.121 | 0.111 | 1.113 |
|  | $W_q$ | 0.018 | 0.011 | 0.011 |
|  | $W_s$ | 0.198 | 0.111 | 0.122 |
| **Moderate Traffic** | $\lambda_{\text{eff}}$ | 0.996 | 1.000 | 0.992 |
|  | $L_q$ | 0.446 | 0.500 | 0.501 |
|  | $L_s$ | 0.978 | 1.000 | 1.003 |
|  | $W_q$ | 0.091 | 0.100 | 0.101 |
|  | $W_s$ | 0.126 | 0.200 | 0.151 |
| **High Traffic** | $\lambda_{\text{eff}}$ | 4.961 | 5.000 | 4.989 |
|  | $L_q$ | 9.428 | 7.551 | 10.269 |
|  | $L_s$ | 10.458 | 8.478 | 11.205 |
|  | $W_q$ | 0.998 | 0.816 | 1.095 |
|  | $W_s$ | 0.934 | 0.915 | 1.189 |
|  | $\lambda_{\text{eff}}$ | 8.955 | 9.260 | 9.379 |

**Table II.** Comparison of quantum simulation and theoretical results for M/M/1 queue metrics across traffic scenarios

These results demonstrate that our quantum method achieves competitive accuracy across different traffic scenarios, with strengths in high-traffic conditions. Although performance varies across metrics, the quantum approach shows promise in estimating queue lengths ($L_q$ and $L_s$) under heavy load, outperforming both theoretical and DES results in these cases. This suggests that our method leverages quantum parallelism effectively to explore multiple queue states simultaneously, potentially offering unique insights into system behavior under congestion. These findings indicate that the quantum approach, although not uniformly superior to classical methods, provides a valuable complementary tool for analyzing complex queueing systems, particularly those with large state spaces where classical methods may struggle.

To explore the effect of the number of qubits on the accuracy of the quantum method, we compared the quantum effective arrival rates ($\lambda_{\text{eff}}$) across different number of qubits. Figure 3 figure presents a comprehensive comparison of the theoretical and quantum effective arrival rates across three traffic scenarios: low, moderate, and high. Each scenario is represented by a separate subplot, illustrating how $\lambda_{\text{eff}}$ varies with the number of qubits used in the quantum simulation, ranging from two to nine qubits. In the low-traffic scenario ($\lambda = 1, \mu = 10$), we observe rapid convergence of the quantum simulation to the theoretical value as the number of qubits increases. The relative error drops significantly from 42.9% with two qubits to 0.49% with three qubits, a reduction of nearly two orders of magnitude. For four or more qubits, the relative error remains consistently below 5%, with the lowest error of 0.098% achieved at eight qubits. This demonstrates the high accuracy of our quantum method in simulating low-traffic conditions, even with a relatively small number of qubits.

The moderate-traffic scenario ($\lambda = 5, \mu = 10$) shows a similar trend of convergence, but with a slightly higher initial discrepancy. The quantum $\lambda_{\text{eff}}$ approaches the theoretical value of 5.00 closely at five qubits, with a relative error of 0.195%. The initial error at two qubits is 4.61%, which is notably lower than in the low-traffic scenario. However, the relative error in this scenario shows more fluctuation, ranging from 0.098% to 6.47% for five to nine qubits, indicating that moderate-traffic conditions may present more challenges to the stability of quantum simulation.

For the high-traffic scenario ($\lambda = 9.5, \mu = 10$), we observe the most pronounced initial difference between theoretical and quantum $\lambda_{\text{eff}}$ values. At two qubits, the relative error is 4.25%, and the convergence is slower than with the other scenarios. The quantum $\lambda_{\text{eff}}$ reaches within 1% of the theoretical value ($\lambda_{\text{eff}} \approx 9.00$) at six qubits, with a relative error of 0.097%. This error remains below 1.3% for higher qubit counts, with the lowest error of 0.293% at nine qubits. These results suggest that high-traffic conditions are the most demanding for our quantum simulation method, requiring more qubits to achieve comparable accuracy.

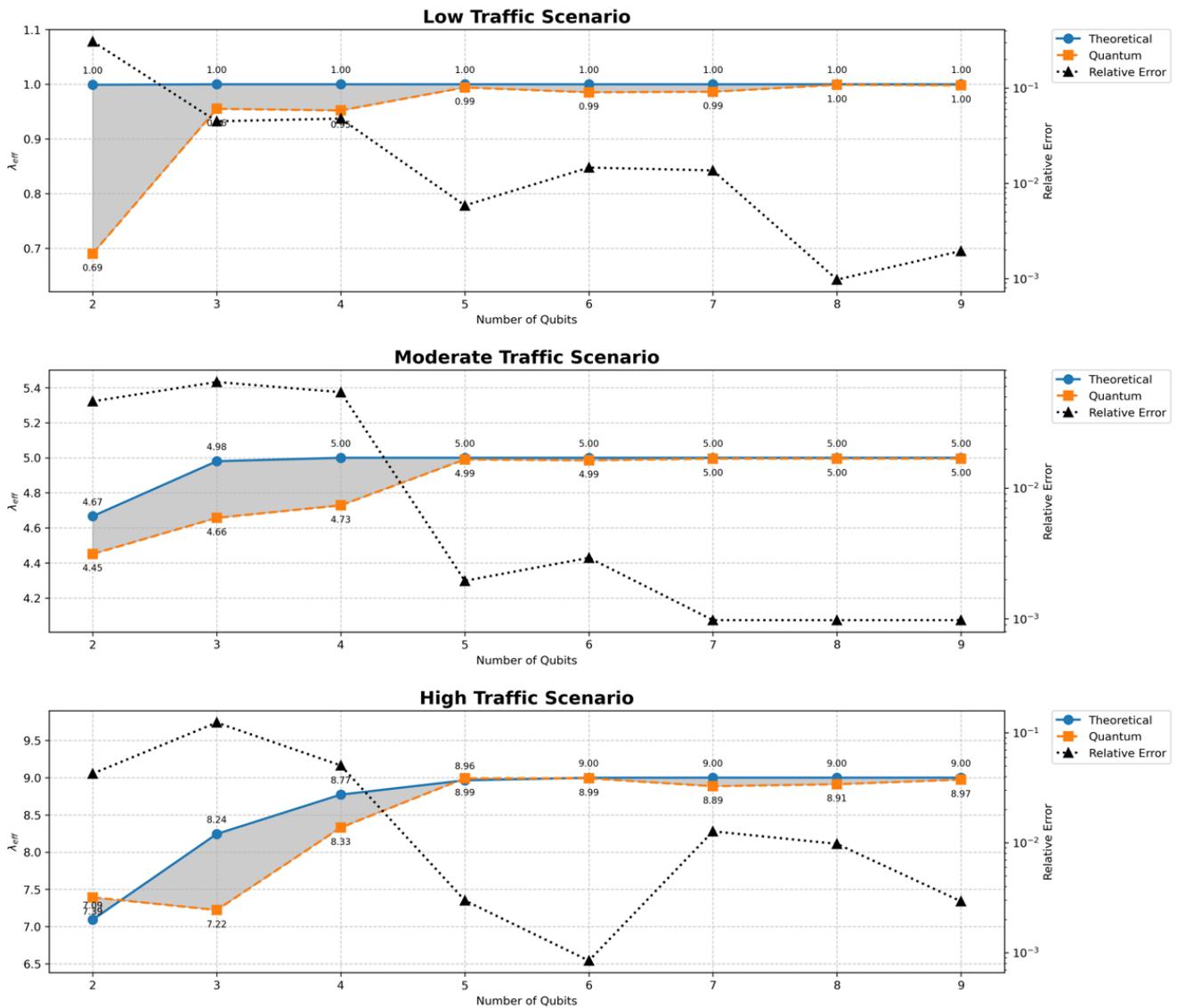

**Figure 3.** Convergence of quantum-simulated $\lambda_{\text{eff}}$ to theoretical values across traffic scenarios and qubit counts

**5.2. Sensitivity analysis**

To assess the robustness of our quantum simulation method and understand the impact of the amplification parameters $\alpha$ and $\beta$, we conducted a comprehensive sensitivity analysis across all three traffic scenarios as described in Section 4.1. Figure 4 presents heatmaps illustrating the relative error in $\lambda_{\text{eff}}$ for each scenario as $\alpha$ and $\beta$ vary.

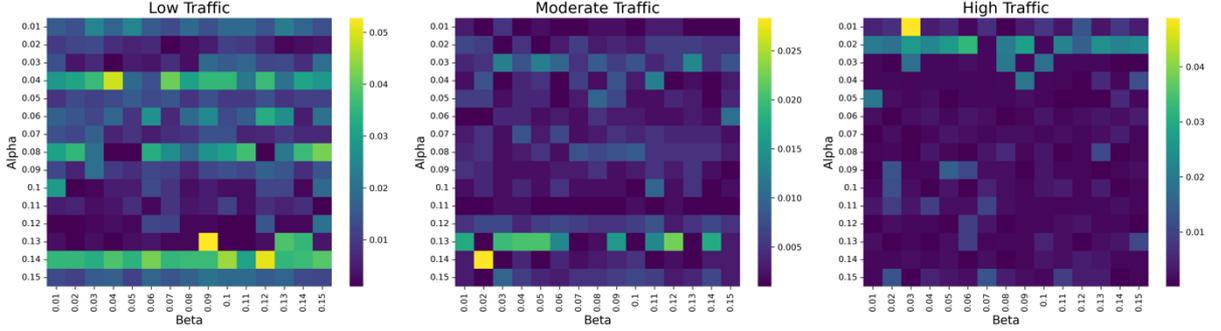

**Figure 4.** Sensitivity analysis of amplification parameters across traffic scenarios

The sensitivity analysis reveals that our quantum simulation method performs well across all traffic scenarios, with the best results obtained in moderate- and high-traffic conditions. High-traffic scenarios show the most consistent performance, with relative errors below 0.01 for a wide range of $\alpha$ and $\beta$ combinations, demonstrating robustness in simulating congested queueing systems.

Low-traffic scenarios exhibit more variation, with relative errors ranging from 0.005 to 0.05, indicating that careful parameter tuning is more critical in these conditions. Optimal performance is achieved with $\alpha$ and $\beta$ values between 0.01 and 0.05. For moderate traffic, optimal values are around $\alpha$ = 0.13–0.15 and $\beta$ = 0.01–0.03, yielding errors below 0.003. High-traffic conditions perform well with $\alpha$ values of 0.08–0.15 and $\beta$ values of 0.10–0.15, achieving errors below 0.002.

The analysis reveals a trade-off in parameter selection, with narrow optimal regions yielding very low relative errors (as low as 0.0001). This highlights the importance of careful parameter tuning for specific traffic conditions. The shifting optimal regions across scenarios demonstrate the method's adaptability to various traffic intensities.

This sensitivity analysis underscores the flexibility and robustness of our quantum simulation approach, providing guidance for parameter selection and highlighting its ability to maintain accuracy across diverse traffic conditions. The consistent performance in moderate- and high-traffic scenarios, often reaching errors as low as 0.0001, suggests that our quantum method offers a promising approach for simulating complex queueing systems and could outperform classical methods in handling challenging high-traffic scenarios.

## 6. DISCUSSION AND CONCLUSION

This study presented a novel quantum method for simulating single-server Markovian (M/M/1) queues, introducing a dynamic amplification approach and custom-parameterized quantum gates for arrival and service processes. The main innovation is the development of a flexible framework that adapts to varying traffic conditions while leveraging quantum computing capabilities, potentially outperforming classical methods in complex queueing scenarios. Here are the main conclusions:

- **High accuracy and rapid convergence across traffic scenarios:** The quantum simulation method demonstrated exceptional accuracy, particularly in high-traffic conditions in which classical methods often struggle. In the high-traffic scenario ($\rho = 0.95$), the method achieved a relative error of only 0.097% for $\lambda_{\text{eff}}$ with six qubits. Moreover, the method exhibited rapid convergence as the number of qubits increased. For instance, in the low-traffic scenario ($\rho = 0.1$), the relative error in $\lambda_{\text{eff}}$ dropped from 42.9% with two qubits to 0.49% with three qubits, a reduction of nearly two orders of magnitude. This combination of accuracy and rapid convergence suggests that the method efficiently leverages quantum resources to simulate complex queueing systems, potentially outperforming classical approaches in challenging scenarios.

- **Robust and adaptive performance through dynamic amplification:** The dynamic amplification approach, implemented through the custom-parameterized quantum gates $A(\lambda, k, \alpha, \Delta t)$ and $S(\mu, k, \beta, \Delta t)$, proved highly effective in adapting to various traffic conditions. The sensitivity analysis revealed that the method performs consistently well across different scenarios and can achieve very low relative errors (as low as 0.0001) when parameters are optimally tuned. The shifting optimal regions for $\alpha$ and $\beta$ across scenarios (e.g., $\alpha = 0.13$–$0.15$, $\beta = 0.01$–$0.03$ for moderate traffic; $\alpha = 0.08$–$0.15$, $\beta = 0.10$–$0.15$ for high traffic) demonstrate the method's flexibility in modeling diverse queueing behaviors.

- **Theoretical foundation bridging quantum computing and queueing theory:** The introduction of custom-parameterized quantum gates, supported by rigorous mathematical proofs (Theorems 1–3), establishes a solid theoretical foundation that bridges classical queueing theory and quantum computing. These gates accurately model Poisson arrivals and exponential service times while incorporating dynamic amplification, providing a novel framework for quantum simulation of queueing systems. This theoretical contribution opens new avenues for exploring the intersection of quantum computing and operations research.

- **Comprehensive simulation of queueing metrics with potential quantum advantage:** The method successfully simulated multiple queueing performance metrics ($L_q, L_s, W_q, W_s, \lambda_{\text{eff}}$) with high accuracy across various traffic intensities. For example, in the high-traffic scenario, the quantum simulation achieved a $\lambda_{\text{eff}}$ of 8.955 compared to the theoretical 8.964. The ability to accurately capture these diverse metrics, combined with the method's efficiency in high-dimensional state spaces, indicates a potential advantage for the quantum approach. This is particularly evident in the simulation of high-traffic scenarios with relatively few qubits (six to nine), for which classical methods typically require significantly more computational resources due to the large state space.

Despite its promising results, this study has several limitations that open avenues for future research. The current method requires careful tuning of amplification parameters, and its performance has only been evaluated for up to nine qubits. Moreover, the impact of quantum noise and errors on simulation results has not been addressed. Future work should focus on developing automated parameter optimization techniques, investigating the method's scalability to larger and more complex queueing systems, and incorporating error mitigation strategies for noisy quantum environments.

To further establish the method's efficacy and broaden its applicability, future research should include comprehensive comparisons with state-of-the-art classical and quantum simulation approaches. Additionally, extending the method to more complex queueing models (e.g., multi-server queues, priority queues, or queueing networks) and exploring its practical applications in fields such as telecommunications, healthcare, or logistics would demonstrate its real-world value. Finally, deeper theoretical investigations could provide new insights into the relationships between quantum computing, queueing theory, and stochastic processes, potentially leading to novel quantum algorithms for operations research problems.